\PassOptionsToPackage{table}{xcolor}
\documentclass[11pt]{article}

\usepackage[most]{tcolorbox}
\usepackage{listings}
\usepackage{xcolor}

\lstdefinestyle{promptstyle}{
    basicstyle=\ttfamily\scriptsize,
    breaklines=true,
    breakatwhitespace=false,
    columns=fullflexible,
    keepspaces=true,
    showstringspaces=false,
    frame=none,
    xleftmargin=0pt,
    xrightmargin=0pt,
    aboveskip=0pt,
    belowskip=0pt
}

\newtcblisting{promptbox}[2][]{
    enhanced,
    breakable,
    listing only,
    listing options={style=promptstyle},
    title={#2},
    fonttitle=\bfseries\small,
    coltitle=black,
    colback=gray!3,
    colframe=black!45,
    colbacktitle=gray!12,
    boxrule=0.4pt,
    arc=1.2mm,
    left=1.2mm,
    right=1.2mm,
    top=1.0mm,
    bottom=1.0mm,
    toptitle=0.8mm,
    bottomtitle=0.8mm,
    #1
}
\usepackage[preprint]{acl}
\usepackage[table]{xcolor}
\definecolor{traceblue}{RGB}{226,239,255}
\definecolor{tracegreen}{RGB}{226,246,232}
\usepackage{times}
\usepackage{latexsym}
\usepackage[T1]{fontenc}
\usepackage[utf8]{inputenc}
\usepackage{microtype}
\usepackage{graphicx}
\usepackage{amsmath}
\usepackage{amssymb}
\usepackage{booktabs}
\usepackage{tabularx}
\usepackage{array}
\usepackage{pifont}
\usepackage{xurl}
\usepackage{xspace}
\usepackage[htt]{hyphenat}
\usepackage{makecell}
\usepackage{inconsolata}
\usepackage{multirow}
\usepackage{enumitem}
\usepackage{tikz}
\DeclareRobustCommand{\gfull}{\tikz[baseline=-0.6ex]\fill (0,0) circle (0.55ex);}
\DeclareRobustCommand{\ghalf}{\tikz[baseline=-0.6ex]{\fill (0,0) -- (90:0.55ex) arc (90:270:0.55ex) -- cycle; \draw[line width=0.4pt] (0,0) circle (0.55ex);}}
\DeclareRobustCommand{\gnone}{\tikz[baseline=-0.6ex]\draw[line width=0.4pt] (0,0) circle (0.55ex);}

\newcommand{\bluehl}[1]{\colorbox{blue!20}{#1}}
\newcommand{\redhl}[1]{\colorbox{red!20}{#1}}
\newcommand{\gatered}[1]{\colorbox{green!15}{\textit{#1}}}
\newcommand{\gateblue}[1]{\colorbox{skyblue!35}{\textit{#1}}}
\newcommand{\gateorange}[1]{\colorbox{orange!15}{\textit{#1}}}
\newcommand{\gategreen}[1]{\colorbox{green!15}{\textit{#1}}}
\newcommand{\codetext}[1]{\texttt{#1}}
\newcommand{\eviAct}{\textsc{EviACT}\xspace}
\newcommand{\notrep}{\textcolor{black!30}{--}} 

\definecolor{skyblue}{RGB}{135,206,235}
\definecolor{orange}{RGB}{255,165,0}

\title{\eviAct: An Evidence-to-Action Framework\\for Agentic Program Repair}

\author{
Qianru Meng$^{1}$ \quad Xiao Zhang$^{2}$ \quad Zhaochun Ren$^{1}$ \quad Joost Visser$^{1}$ \\
$^{1}$LIACS, Leiden University, Leiden, Netherlands \\
$^{2}$CLCG, University of Groningen, Groningen, Netherlands \\
\texttt{mengqr@vuw.leidenuniv.nl}, \texttt{xiao.zhang@rug.nl} \\
\texttt{\{z.ren,j.m.w.visser\}@liacs.leidenuniv.nl}
}
\begin{document}
\maketitle

\begin{abstract}
\looseness=-1 LLM-based agents have moved automated program repair (APR) from fixed-context patch generation to interactive repository-level repair.
Existing agentic APR systems observe execution evidence such as failing tests and compiler errors, yet still struggle to use it to guide localization, patch generation, and validation.
We propose \eviAct{} (\emph{\textbf{Evi}dence-to-\textbf{Act}ion}), an agentic APR framework that coordinates three evidence-driven guardrails across repair stages.
The retrieval scaffold grounds repair context, the compile gate filters invalid edits, and the test-driven gate checks target-test recovery before full regression.
Across four benchmarks, \eviAct{} improves resolve rate over the strongest published GPT-4o baselines by 6.0--17.5 percentage points at 72.9--90.5\% lower per-instance API cost on Lite and Live, the two benchmarks where baseline costs are reported. With the same DeepSeek-V4-Pro backbone and budgets, it outperforms the strongest rerun baseline, mini-SWE-agent, by 10.0--19.0 percentage points on SWE-bench.
Ablations and diagnostics trace these gains to the coordinated evidence-to-action chain, which makes agentic APR both more effective and more efficient.
\end{abstract}

\section{Introduction}

Large language models (LLMs) have enabled automated program repair (APR) systems to fix real bugs in software repositories~\citep{yang2024swe-agent,zhang2024autocoderover}. Traditional APR techniques search a bounded patch space under test oracles or learn repair patterns from historical fixes, and operate within a fixed repair context~\citep{weimer2009genprog,long2016prophet,xia2022alphaRepair}. Given an issue and a repository, a system must locate the faulty code, generate a patch, and validate that the patch fixes the bug without breaking existing behavior. These steps are tightly coupled. Issues describe symptoms in terms that rarely match the implementation, so fault localization often selects the wrong code, and every subsequent step inherits this error~\citep{reddy2026swerank}. Validation, in turn, produces the evidence needed to correct it, as a compiler error or a failing target test indicates whether the patch or the selected location is at fault.

\begin{figure}[t]
    \centering
    \includegraphics[width=\columnwidth]{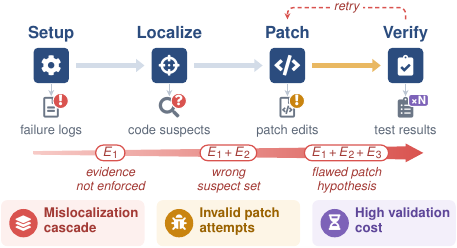}
    \caption{Problems in workflow-based agentic APR. Errors $E_1$--$E_3$ accumulate across stages in a mislocalization cascade, invalid patch attempts reach validation, and repeated regression raises validation cost.}
    \label{fig:problem_statement}
\end{figure}

Interaction-based agents perform these steps through open-ended exploration. SWE-agent and OpenHands give the LLM tools to inspect code, edit files, and run tests, and leave it to the model to decide what to inspect, which tool to call, and when to change course~\citep{yang2024swe-agent,wang2024openhands}. Each decision consumes tokens, and coarse-grained execution feedback offers little guidance on where to look next~\citep{reddy2026swerank,xiang2026framepilot}.

\looseness=-1 Workflow-based approaches reduce this planning burden by decomposing repair into predefined stages. Agentless separates localization, patch generation, and validation in a simple and economical pipeline~\citep{xia2024agentless}. TDFlow assigns test-driven subtasks to specialized agents, which narrows the context each agent handles~\citep{lee2025tdflow}. However, decomposition does not remove the dependencies between stages. Patch generation still works on the code that localization selected, and higher localization accuracy coincides with higher resolve rates. When this code is wrong, a rejected patch does not reveal where the fix belongs, so failure feedback must also guide localization.

Both paradigms collect this execution evidence, and some already enforce checks on it. Agentless filters and reranks candidate patches with reproduction and regression tests, and TDFlow organizes development around tests that must pass. These checks determine whether a candidate patch is accepted. In Agentless, they do not decide what the next stage works on, so a rejected patch is regenerated for the same suspect code and the agent keeps retrying under the same incorrect hypothesis. Figure~\ref{fig:problem_statement} illustrates this cascade, and in our case study of Appendix~\ref{app:case}, a variant without guardrails makes 19 failed patch attempts on a mislocalized file until its budget runs out. We refer to this disconnect between execution evidence and the next action as the \emph{evidence-to-action gap}, which this work sets out to close.

\looseness=-1 We propose \eviAct{} (\emph{\textbf{Evi}dence-to-\textbf{Act}ion}), an APR framework that closes this gap with three evidence-driven guardrails at the stage boundaries of a Setup--Localize--Patch--Verify pipeline. Before patching, the \emph{retrieval scaffold} combines the failure log of the reproduced bug (RED) with code structure to select a small set of suspicious code spans, the \emph{suspect set}. Before validation, the \emph{compile gate} checks that a patch applies and compiles and returns diagnostics for refinement when it does not. Before full regression, the \emph{test-driven (TD) gate} requires the target tests to pass (GREEN). When repeated attempts fail these tests and budget remains, \eviAct{} returns to localization and reconsiders the suspect set. Execution evidence thus decides where to edit, whether to refine a patch, and when to run full validation. A guardrail therefore differs from a filter in how its outcome is used. In \eviAct{}, a check decides whether the next stage runs and what it acts on, as when a failing target test sends the repair back to localization. This makes enforcement a design choice that we evaluate in isolation by running the same checks as advisory feedback.

We evaluate \eviAct{} on Defects4J~2.0, SWE-bench Verified, SWE-bench Lite, and SWE-bench Live with five LLM backbones. With the same DeepSeek-V4-Pro backbone and repair budget, \eviAct{} outperforms mini-SWE-agent by 10.0--19.0 percentage points on the three SWE-bench sets at a per-instance cost of at most \$0.08. With GPT-4o, it exceeds the strongest published GPT-4o baseline on every benchmark, by 6.0--17.5 points. Ablation studies show that each guardrail contributes, and enforcing the same checks yields 8.3 more points than advisory feedback does.

\noindent Our contributions are as follows.
\begin{itemize}[leftmargin=*,nosep]
    \item We identify the evidence-to-action gap in agentic APR and formulate guardrails as evidence checks that decide whether the next stage runs and what it acts on.
    \item We propose \eviAct{}, which coordinates three guardrails to guide localization, patch refinement, and validation, and returns to localization when target tests keep failing.
    \item We evaluate \eviAct{} on four benchmarks with five backbones, isolate each guardrail and the effect of enforcement through ablation, and manually analyze the residual failures and the correctness of resolved patches.
\end{itemize}

\begin{figure*}[t]
    \centering
    \includegraphics[width=\textwidth]{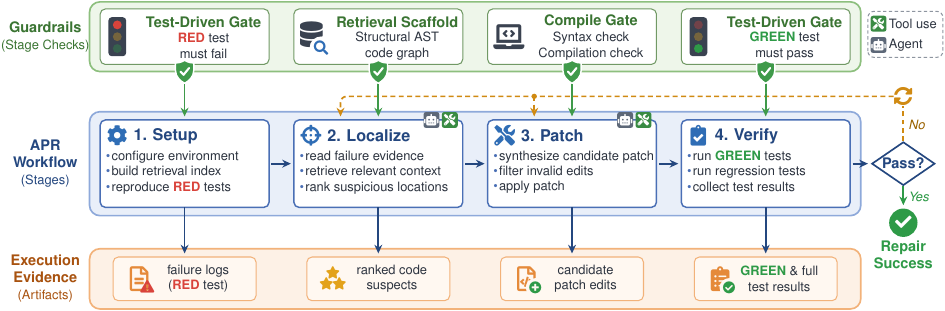}
    \caption{Overview of \eviAct{}. The retrieval scaffold derives code suspects from RED failure logs, the compile gate filters invalid patch candidates, and the TD gate requires GREEN target tests before full regression. The Localize and Patch agents call tools to inspect suspects and refine edits.}
    \label{fig:pipeline}
\end{figure*}

\section{Related Work}
\label{sec:related_work}

Classical APR searches bounded patch spaces under test oracles~\citep{weimer2009genprog,le2012genprog,nguyen2013semfix,mechtaev2016angelix,long2016prophet}, and learning-based repair learns repair patterns from historical fixes~\citep{wong2021varfix,parasaram2021trident,zhang2023survey}.
LLM-based agents move repair to the repository level, where they inspect code, invoke tools, edit files, and run tests~\citep{yang2024swe-agent,wang2024openhands,zhang2024autocoderover,arora2024masai,xia2024agentless,chen2024coder,li2025patchpilot,yang2025survey,yu2025patchagent}.
Interaction-based agents such as SWE-agent, OpenHands, AutoCodeRover, MASAI, and mini-SWE-agent let the model drive exploration through flexible tool interfaces~\citep{yang2024swe-agent,wang2024openhands,zhang2024autocoderover,arora2024masai,miniswe2025}. Workflow-based systems such as Agentless, PatchPilot, and CoDeR separate localization, patch construction, and validation into explicit stages~\citep{xia2024agentless,li2025patchpilot,chen2024coder}.
Recent agents further improve repair through self-evolving scaffolds or experience reused across instances~\citep{xia2025liveswe,mu2026experepair}.
\eviAct{} follows the workflow design and focuses on how evidence controls stage transitions.

Existing systems use execution evidence within individual stages.
Localization combines repository structure, learned retrievers, dependency relations, or spectrum-based signals~\citep{xia2024agentless,zhang2024autocoderover,wei2025swefixer,ma2025lingma,tang2025synfix}. Patch admission filters or ranks candidates through executability checks, semantic review, specification reasoning, adversarial intent reasoning, or multi-candidate selection~\citep{li2025patchpilot,pabba2025semagent,tang2025copatcher,ruan2025specrover,liu2024marscode,ye2025adverintent}. TDFlow organizes software engineering around RED--GREEN--REFACTOR feedback~\citep{lee2025tdflow}.
Work on patch overfitting and safety shows that test-passing patches are not always correct~\citep{kim2025logs,hu2025sok,cambronero2025abstain}.
In these systems, a check filters candidates within a stage or provides context for the model.
\eviAct{} uses failure logs, compiler diagnostics, and target-test outcomes as guardrails that decide whether and on what the next stage runs. Appendix~\ref{app:positioning} compares the mechanisms at each stage boundary.

\section{Methodology}
\label{sec:method}

\subsection{Overview}

As shown in Figure~\ref{fig:pipeline}, \eviAct follows the common staged structure of workflow-based APR systems, using Setup, Localize, Patch, and Verify as its execution backbone. It differs in using execution evidence as a control signal at stage transitions, where repair trajectories tend to drift.

(1) Setup initializes the repair environment, constructs a structure-preserving retrieval index, and reproduces the target failure to obtain RED evidence. (2) Localize converts this evidence into failure-grounded queries, expands over structural code relations, and ranks suspicious locations. (3) Patch synthesizes candidate edits over the localized suspect set and filters invalid candidates before validation. (4) Verify reruns the target tests before full regression and accepts a patch only when both validation stages pass.

We formalize each guardrail at a stage boundary as a triple $(e, c, a)$ of an evidence artifact $e$ produced upstream, an admissibility check $c(e)$, and a constraint $a$ on the downstream action space that applies only when $c(e)$ holds. When the check fails, its outcome returns upstream as feedback. A transition is allowed only when its guardrail check holds, so no stage acts on an artifact that its guardrail has rejected. Mechanisms in prior systems are special cases of this triple. An advisory check returns $c(e)$ to the model but imposes no constraint $a$, and a stage-local filter applies $c(e)$ to candidates within a stage but leaves the action space of the next stage unconstrained. Three such guardrails control the workflow. The \emph{retrieval scaffold} bridges Setup and Localize by mapping RED evidence to a focused suspect set over AST spans and code-graph relations. The \emph{compile gate} operates in Patch by rejecting malformed, unappliable, or non-compilable edits before test validation. The \emph{TD gate} spans Setup and Verify. RED reproduction anchors localization, and GREEN validation requires the target tests to pass before regression.

\subsection{Setup}

Setup produces two artifacts used throughout repair, a structure-preserving retrieval index $\mathcal{I}$ and RED failure evidence from the target tests. It first checks out the buggy revision of the repository into a clean workspace $s_0$.

The retrieval index preserves the structure of the code and is built in three steps. First, \eviAct parses source files into ASTs using Tree-sitter\footnote{\url{https://tree-sitter.github.io/tree-sitter/}} and extracts structural spans at file, class, method, and function granularity. Each span stores its file path, line range, symbol name, signature when available, and local code content. Second, following the cascaded-compression design of ABCoder,\footnote{\url{https://github.com/cloudwego/abcoder}} \eviAct performs offline, LLM-free deterministic compression. Lower-level spans retain executable code, and higher-level spans store structural descriptors derived from child symbols, signatures, imports, and containment metadata. Third, \eviAct extracts containment, call, import, and inheritance edges between spans, forming a lightweight code graph for localization.

For a repository $R$, \eviAct constructs
\begin{equation}
\begin{aligned}
\mathcal{S}_0 &= \textsc{Spans}(\textsc{ASTParse}(R)),\\
\mathcal{S} &= \textsc{Compress}(\mathcal{S}_0),\\
\mathcal{I} &= (\mathcal{S}, \mathcal{E}),
\end{aligned}
\label{eq:index_construction}
\end{equation}
where $\mathcal{S}$ is the set of compressed structural spans and $\mathcal{E}$ holds the code-graph edges.

Setup also executes the RED phase of the TD gate by reproducing the target failure in the clean workspace. On Defects4J~2.0 it runs the benchmark-provided triggering tests, and on SWE-bench it runs reproduction tests that the LLM generates from the issue, following Agentless~\citep{xia2024agentless}. The resulting RED evidence includes the failing test identifier, error message, stack trace, and assertion failure when available. This RED evidence anchors localization.

\subsection{Localize}

Given RED evidence and the retrieval index $\mathcal{I}=(\mathcal{S},\mathcal{E})$, the retrieval scaffold constructs a focused suspect set for patch generation.

\eviAct retrieves code deterministically over the code graph. It first maps RED evidence to seed spans by matching failure-grounded signals from the failing test name, stack trace, error message, assertion content, and available file or line references against structural span fields, including file path, line range, symbol name, signature, local code content, and structural descriptor. It then expands from these seed spans over the code graph. Under a bounded search budget, the scaffold traverses containment, call, import, and inheritance edges to recover code regions structurally related to the directly observed failure location.

Finally, \eviAct ranks candidate spans using deterministic evidence alignment. For each candidate span $s$, the ranking key is
\begin{equation}
\rho(s)=
\bigl(
m_{\textsc{sym}}(s),\!
m_{\textsc{loc}}(s),\!
-d_G(s),\!
r_G(s),\!
-|s|
\bigr),
\label{eq:localization_rank}
\end{equation}
where $m_{\textsc{sym}}(s)$ denotes exact trace-symbol match, $m_{\textsc{loc}}(s)$ denotes file or line proximity when available, $d_G(s)$ is the graph distance from RED-matched seeds, $r_G(s)$ measures relation support in the expanded subgraph, and $|s|$ is the span length. Candidates are sorted in descending lexicographic order by $\rho(s)$, and the top-$K$ spans form the suspect set $\mathcal{U}_K$, with $K=3$ as a fixed compact default. Appendix~\ref{app:config_rationale} discusses the rationale.

\subsection{Patch}

Conditioned on the suspect set $\mathcal{U}_K$ and its supporting evidence, the patch agent generates candidate diffs through bounded tool use. At each attempt, it receives the localized working set, RED evidence, and current diagnostic context. The agent may inspect nearby code, but patch generation stays within the suspect set until re-localization.

Before test validation, \eviAct applies the compile gate to each candidate patch $p$,
\begin{equation}
g_{\textsc{cmp}}(p)
=
I_{\textsc{diff}}(p)
\cdot
I_{\textsc{apply}}(p)
\cdot
I_{\textsc{comp}}(s_0 \oplus p).
\label{eq:compile_gate}
\end{equation}
Here, $I_{\textsc{diff}}(p)$, $I_{\textsc{apply}}(p)$, and $I_{\textsc{comp}}(s_0 \oplus p)$ indicate whether the patch is a well-formed diff, applies cleanly, and passes syntax or build checks, respectively. Thus, $g_{\textsc{cmp}}(p)=1$ only when the patch passes all three checks.

When the compile gate fails, \eviAct returns structured diagnostics, including the failure type, affected file or hunk when available, and relevant syntax or build output, as refinement context for the next patch attempt. Only candidates that pass the compile gate proceed to verification.

\subsection{Verify}

The Verify stage applies the GREEN phase of the TD gate before full regression. For a compiled patch $p$, \eviAct first reruns the target tests on the patched workspace $s_0 \oplus p$. If these GREEN tests fail, the resulting failure evidence is returned to the patch agent for refinement. Repeated GREEN failures under the same suspect set trigger re-localization when budget remains, which happens in about 15\% of SWE-bench Lite instances with DeepSeek-V4-Pro. Of these, 24\% are then resolved, adding about 3.6 pp on Lite.

Only GREEN-passing patches proceed to full regression validation. This ordering separates target-failure repair from regression preservation. Formally, the final acceptance condition is
\begin{equation}
g_{\textsc{td}}(p)
=
I_{GREEN}(s_0 \oplus p)
\cdot
I_{\textsc{reg}}(s_0 \oplus p),
\label{eq:td_gate}
\end{equation}
where $I_{GREEN}$ indicates whether the target tests pass and $I_{\textsc{reg}}$ indicates whether the full regression suite passes. A patch is accepted only when $g_{\textsc{td}}(p)=1$.

\section{Experimental Setup}
\label{sec:setup}

\begin{table}[t]
\centering
\small
\setlength{\tabcolsep}{4pt}
\renewcommand{\arraystretch}{1.08}
\begin{tabularx}{\linewidth}{@{}lccc@{}}
\toprule
\textbf{Dataset} & \textbf{Language} & \textbf{Number} & \textbf{Coverage} \\
\midrule
Defects4J~2.0 & Java & 835 & 17 \\
SWE-bench Verified & Python & 500 & 12  \\
SWE-bench Lite & Python & 300 & 11 \\
SWE-bench Live & Python & 100 & 55 \\
\bottomrule
\end{tabularx}
\caption{Benchmark settings used in our evaluation, including the programming language, number of instances, and repository/project coverage.}
\label{tab:benchmarks}
\end{table}

\subsection{Research Questions}
We address three research questions.
RQ1: How does \eviAct{} perform against existing agentic repair systems across benchmarks and backbones, and at what cost?
RQ2: How does each guardrail contribute, and does enforcing the guardrails matter beyond exposing their evidence?
RQ3: Where does \eviAct{} still fail, and are its resolved patches semantically correct?

\begin{table*}[t]
\centering
\small
\setlength{\heavyrulewidth}{0.7pt}
\setlength{\lightrulewidth}{0.4pt}
\setlength{\cmidrulewidth}{0.4pt}
\renewcommand{\arraystretch}{1.02}
\setlength{\tabcolsep}{4pt}
\begin{tabularx}{\textwidth}{@{\hspace{4pt}}ll*{8}{>{\centering\arraybackslash}X}@{\hspace{4pt}}}
\toprule
& & \multicolumn{2}{c}{\makebox[0pt]{Defects4J 2.0}} & \multicolumn{2}{c}{\makebox[0pt]{SWE-bench Verified}} & \multicolumn{2}{c}{\makebox[0pt]{SWE-bench Lite}} & \multicolumn{2}{c}{\makebox[0pt]{SWE-bench Live}} \\
\cmidrule(lr){3-4}\cmidrule(lr){5-6}\cmidrule(lr){7-8}\cmidrule(lr){9-10}
Method & Backbone & Res.$\uparrow$ & Cost$\downarrow$ & Res.$\uparrow$ & Cost$\downarrow$ & Res.$\uparrow$ & Cost$\downarrow$ & Res.$\uparrow$ & Cost$\downarrow$ \\
\midrule
RepairAgent & GPT-3.5-Turbo & 22.3 & 0.14 & \notrep & \notrep & \notrep & \notrep & \notrep & \notrep \\
AdverIntent & GPT-4o & 21.6 & \notrep & \notrep & \notrep & \notrep & \notrep & \notrep & \notrep \\
SWE-agent & GPT-4o & \notrep & \notrep & 23.2 & \notrep & 18.3 & 2.53 & 10.0 & 2.42 \\
OpenHands & GPT-4o & \notrep & \notrep & \notrep & \notrep & 22.0 & 1.72 & 8.0 & 1.89 \\
Agentless & GPT-4o & \notrep & \notrep & 38.8 & \notrep & 32.0 & 0.70 & 8.0 & \notrep \\
Agentless$^{*}$ & DeepSeek-V4-Pro & \notrep & \notrep & 48.8 & 0.19 & 42.0 & 0.13 & 9.0 & 0.21 \\
mini-SWE-agent$^{*}$ & DeepSeek-V4-Pro & \notrep & \notrep & 65.6 & 0.05 & 66.3 & 0.06 & 25.0 & 0.08 \\
\midrule
\multirow{5}{*}{\eviAct{}} & GPT-4o & 29.0 & 0.17 & 45.6 & 0.20 & 49.5 & 0.19 & 16.0 & 0.23 \\
& Gemini-3.8-Flash & 69.3 & 0.31 & 68.4 & 0.28 & 75.2 & 0.23 & 36.0 & 0.36 \\
& DeepSeek-V4-Pro & 70.9 & 0.08 & 75.6 & 0.08 & 81.0 & 0.01 & 44.0 & 0.07 \\
& GPT-5.6 & \underline{74.0} & 0.16 & \underline{79.0} & 0.16 & \underline{93.7} & 0.10 & \underline{49.0} & 0.25 \\
& Claude Opus 5.5 & \textbf{77.8} & 0.62 & \textbf{82.6} & 0.52 & \textbf{95.7} & 0.23 & \textbf{53.0} & 0.80 \\
\bottomrule
\end{tabularx}
\caption{Results on Defects4J~2.0 (835 bugs) and SWE-bench Verified (500), Lite (300), and Live (100), reported as resolve rate (Res., \%) and average per-instance API cost (\$). $^{*}$ marks reruns with the same instances, budgets, and cost accounting as \eviAct{}, averaged over three runs. The remaining baselines are published results, and -- marks results that were not reported. Defects4J baselines assume perfect fault localization. \textbf{Bold} and \underline{underline} mark the best and second-best resolve rate.}
\label{tab:main_results}
\end{table*}

\subsection{Datasets and Metrics}
\label{sec:datasets}

We evaluate \eviAct on Defects4J~2.0~\cite{just2014defects4j}, SWE-bench Verified~\cite{chowdhury2024swebenchverified}, SWE-bench Lite~\cite{jimenez2024swebench}, and SWE-bench Live~\cite{zhang2026swe}.
Since SWE-bench Live is designed to reduce benchmark leakage from public training and evaluation artifacts, we randomly sample 100 instances for evaluation, covering 55 repositories.
Defects4J~2.0 contains real Java bugs with triggering tests, and the three SWE-bench sets contain Python GitHub issues validated by repository tests. Table~\ref{tab:benchmarks} summarizes their language, size, and repository coverage.

We report resolve rate as the primary effectiveness metric, defined as the percentage of instances for which the accepted patch passes the benchmark validation suite.
We report average per-instance API cost in USD as the primary efficiency metric.
For ablation and error analysis we also report Hit@3, \textsc{CompFail}, \textsc{ValFail}, tokens, runtime, and tool calls, defined in Appendix~\ref{app:metrics}.

\subsection{Baselines and Models}

We compare with RepairAgent~\cite{repairagent_ref} and AdverIntent~\cite{ye2025adverintent} on Defects4J, and SWE-agent~\cite{yang2024swe-agent}, Agentless~\cite{xia2024agentless}, and OpenHands~\cite{wang2024openhands} on SWE-bench. These systems cover both interaction-based and workflow-based designs, and we report the results in their papers.

Our primary framework comparison reruns Agentless and mini-SWE-agent~\cite{miniswe2025} with DeepSeek-V4-Pro on the same instances, with the same 45-minute and 56-tool-call budgets, execution environment, and cost accounting as \eviAct. Results are averaged over three runs. Each baseline keeps its own repair workflow, as detailed in Appendix~\ref{app:baseline_reruns}.

\looseness=-1 We evaluate GPT-4o~\cite{openai2024gpt4o}, Gemini-3.8-Flash~\cite{google2026gemini38flash}, DeepSeek-V4-Pro~\cite{deepseek2026deepseekv4pro}, GPT-5.6~\cite{openai2026gpt56}, and Claude Opus 5.5~\cite{anthropic2026opus55}. All ablations and manual analyses use \mbox{DeepSeek-V4-Pro} as the backbone.

\subsection{Implementation}
\eviAct{} is implemented in Python and interfaces with benchmark-specific repository, build, and test environments.
Each run stops when a patch is accepted, after 45 minutes, or after 56 tool calls, allocated as 36 localization calls and 20 patching calls. These values are fixed for all benchmarks and backbones, and every rerun baseline receives the same 56-call budget, so no comparison depends on tuning them. Appendix~\ref{app:config_rationale} discusses the allocation.
All LLMs are accessed through vendor APIs. We set temperature and top-$p$ to 1.0 where the provider exposes them and use provider defaults otherwise. Prompt details are provided in Appendix~\ref{app:prompts}.

\section{Results and Analysis}
\label{sec:results}
We first compare \eviAct{} with existing systems (RQ1), then ablate its guardrails (RQ2), and finally analyze its failures and resolved patches (RQ3).

\subsection{Main Results (RQ1)}
\label{sec:main_results}
Table~\ref{tab:main_results} reports resolve rate and cost on the four benchmarks. Because published baselines differ in backbone and budget, we first compare against reruns under identical conditions.

\eviAct{} outperforms existing agentic repair systems under the same backbone and budget. With DeepSeek-V4-Pro, it improves over mini-SWE-agent by 10.0--19.0 percentage points (pp) and over Agentless by 26.8--39.0 pp on Verified, Lite, and Live, and every gain is significant under a two-sided exact McNemar test with $p<0.001$. On Lite, for instance, the paired bootstrap 95\% confidence interval of the gain over mini-SWE-agent is [11.0, 18.4] pp. The margin over mini-SWE-agent is largest on Live, where mini-SWE-agent resolves only 25.0\% of instances. With GPT-4o, \eviAct{} also exceeds the strongest published GPT-4o baseline on every benchmark, by 6.0 pp on Live and up to 17.5 pp on Lite.

\eviAct{} achieves these gains at comparable or lower cost. With DeepSeek-V4-Pro, it costs \$0.01--\$0.08 per instance, 57.9--92.3\% less than Agentless, and it is cheaper than mini-SWE-agent on Lite and Live and \$0.03 more on Verified. On Lite, it uses 145K tokens and 348.5s per instance, compared with 1,312K and 433.6s for mini-SWE-agent and 1,549K and 1943.5s for Agentless. With GPT-4o, its cost on Lite and Live is 72.9--90.5\% lower than that of the strongest published baselines.

\begin{figure}[t]
    \centering
    \includegraphics{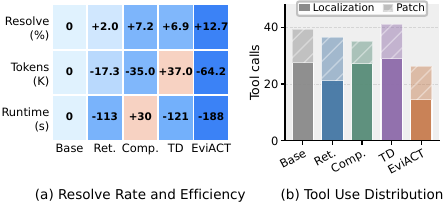}
    \caption{Overall component ablation of \eviAct. 
    (a) Changes in resolve rate, token usage, and runtime relative to the no-guardrail baseline. 
    (b) Tool-call distribution across the Localize and Patch stages.}
    \label{fig:ablation}
\end{figure}

\begin{figure*}[t]
    \centering
    \includegraphics{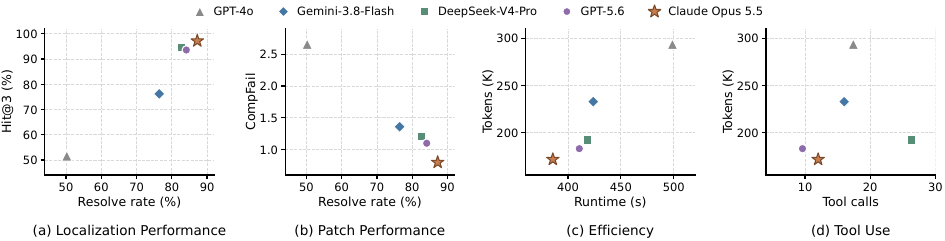}
    \caption{Cross-backbone diagnostics of \eviAct{} on SWE-bench Lite (300 instances), averaged per instance. (a) Hit@3 against resolve rate. (b) \textsc{CompFail}, the number of candidates rejected by the compile gate, against resolve rate. (c) Tokens against runtime. (d) Tokens against tool calls.}
    \label{fig:comparison_llms}
\end{figure*}

The gains grow with backbone capability. Claude Opus 5.5 achieves the highest resolve rate on all four benchmarks, including 82.6\% on Verified and 95.7\% on Lite, while DeepSeek-V4-Pro costs \$0.01--\$0.08 per instance, GPT-5.6 \$0.10--\$0.25, and Opus \$0.23--\$0.80. On Lite, DeepSeek-V4-Pro is the cheapest at \$0.01, and GPT-5.6 resolves 12.7 pp more instances at \$0.10. Figure~\ref{fig:comparison_llms} shows that on Lite, compile rejections and runtime fall steadily as the resolve rate rises, and Hit@3 rises from 51.6\% with GPT-4o to above 93\% for the three strongest backbones. Opus reaches 97.2\% Hit@3 with the shortest runtime, and compile rejections fall from 2.66 with GPT-4o to 0.8 with Opus. Every backbone resolves more on Lite than on Verified because Lite keeps only issues fixed within one file and at most three hunks~\citep{jimenez2024swebench}, which favors a small suspect set. Live is the hardest setting for every backbone, and the strongest one resolves 53.0\% of its recent issues.

\subsection{Ablation Studies (RQ2)}
\label{sec:ablation}
\label{sec:diagnostic}
To identify which guardrail drives these gains and whether enforcement matters, we compare full \eviAct{} with four variants on all 300 Lite instances using DeepSeek-V4-Pro. The no-guardrail variant, labeled Base or Baseline in the figures, removes all three guardrails, and +Retrieval, +Compile, and +TD each add one of them back. The advisory variant runs every check but only returns its outcome to the agent. It still produces the suspect set, compile diagnostics, and GREEN outcomes, yet never blocks a stage transition, so it differs from \eviAct{} in enforcement alone.

Each guardrail improves repair, and the full system yields the largest gain. As Figure~\ref{fig:ablation}a shows, retrieval, the compile gate, and the TD gate add 2.0, 7.2, and 6.8 pp over the no-guardrail variant, averaged over three runs. Full \eviAct{} adds 12.7 pp and reaches the 81.0\% in Table~\ref{tab:main_results}, while using 64.2K fewer tokens, 188s less runtime, and the fewest tool calls, 26.3 against 39.3.

\begin{figure}[t]
    \centering
    \includegraphics{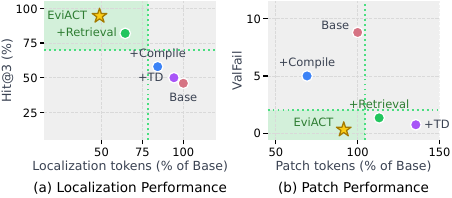}
    \caption{Stage-level ablation analysis of \eviAct. 
    (a) Localization performance, comparing file-level Hit@3 against localization tokens relative to Base. 
    (b) Patch performance, comparing validation failures against patch tokens relative to Base.}
    \label{fig:ablation_details}
\end{figure}

Enforcement matters beyond the evidence itself. The advisory variant observes the same evidence yet resolves 72.7\% of instances, 8.3 pp below \eviAct{} (two-sided exact McNemar $p<0.001$, 95\% confidence interval of the gap [5.7, 11.1] pp), and costs more per instance (\$0.034 vs.\ \$0.01), so enforcement improves both resolve rate and cost. In terms of the triple $(e, c, a)$ of Section~\ref{sec:method}, the advisory variant keeps $e$ and $c$ and drops $a$, so the 8.3 pp measure the constraint on the next stage. Under majority voting over runs, 42 instances are resolved only with enforced guardrails and 11 only with advisory ones.

Each guardrail acts at the stage it guards. Figure~\ref{fig:ablation_details}a shows that retrieval brings the largest localization gain, raising file-level Hit@3 from 46.0\% to 71.0\% while cutting localization tokens by 35.5\%. The compile and TD gates act downstream of localization and raise Hit@3 only to 58.0\% and 50.0\%, so retrieval is the prerequisite for accurate localization. Without the retrieval scaffold, the patch agent is not bound to the top three files and often reaches the faulty file later at a higher cost. These variants therefore resolve more instances than their Hit@3 suggests, and retrieval alone adds only 2.0 pp while cutting 17.3K tokens and 113s per instance. The guardrails are complementary. Retrieval alone reaches 71.0\% Hit@3 and either gate alone at most 58.0\%, yet together they reach 94.6\% with 51.2\% fewer localization tokens, because the RED failure log produced by the TD gate becomes the retrieval query, which the +Retrieval variant lacks. Full \eviAct{} also saves 64.2K tokens per instance, against 15.3K for the three single-guardrail variants combined. In the patch stage, Figure~\ref{fig:ablation_details}b shows that every guardrail lowers validation failures from 8.80, with the TD gate alone reaching 0.74 and full \eviAct{} reaching 0.32. Retrieval alone reaches 1.34 because fewer patches target incorrect locations, and the compile gate cuts patch tokens by 30.9\%.

\begin{figure}[t]
    \centering
    \includegraphics{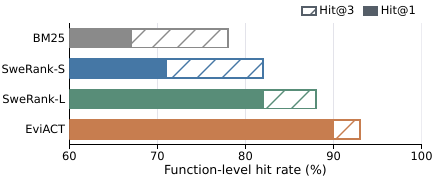}
    \caption{Function-level Hit@1 (solid) and Hit@3 (full hatched bar) of retrieval backends on SWE-bench Lite (300 instances) with DeepSeek-V4-Pro. SweRank-S/L denote SweRankEmbed-Small/Large, and all backends run within full \eviAct{}.}
    \label{fig:retrieval}
\end{figure}

Evidence-driven retrieval outperforms lexical and learned retrievers. Figure~\ref{fig:retrieval} compares the retrieval scaffold under the same localization controller with BM25, which matches issue text lexically, and with SweRankEmbed, which ranks code with learned embeddings. Starting from the stack frames and symbols in the RED log, the scaffold reaches 90.0\% function-level Hit@1 and 93.0\% Hit@3, exceeding SweRankEmbed-Large by 8.0 and 5.0 pp and BM25 by 23.0 and 15.0 pp.

\begin{table}[t]
  \centering
  \small
  \setlength{\tabcolsep}{3pt}
  \renewcommand{\arraystretch}{1.08}
  \begin{tabularx}{\columnwidth}{@{}Xrrrr@{}}
    \toprule
    \textbf{Dataset} & \textbf{Misloc.} & \textbf{Mis-edit} & \textbf{Partial} & \textbf{Timeout} \\
    \midrule
    Defects4J 2.0        & 48.0 & 38.0 & 12.0 & 2.0 \\
    SWE-bench Verified   & 42.0 & 42.0 & 14.0 & 2.0 \\
    SWE-bench Lite       & 46.0 & 36.0 & 14.0 & 4.0 \\
    SWE-bench Live       & 44.0 & 40.0 & 14.0 & 2.0 \\
    \midrule
    Overall (avg.)       & 45.0 & 39.0 & 13.5 & 2.5 \\
    \bottomrule
  \end{tabularx}
\caption{Failure modes of \eviAct{} with DeepSeek-V4-Pro (\%). Misloc.\ is a missed repair context, Mis-edit a wrong edit in the correct context, Partial an incomplete repair of the observed failure, and Timeout a run that exhausts its time or tool budget.}
\label{tab:failure_modes}
\end{table}

\begin{table}[t]
\centering
\small
\setlength{\tabcolsep}{4pt}
\renewcommand{\arraystretch}{1.08}
\begin{tabular}{lrrr}
\toprule
\textbf{Dataset} & \textbf{Correct} & \textbf{Overfitting} & \textbf{Unknown} \\
\midrule
Defects4J 2.0        & 80.0 & 12.0 & 8.0 \\
SWE-bench Verified   & 76.0 & 12.0 & 12.0 \\
SWE-bench Lite       & 80.0 & 8.0 & 12.0 \\
SWE-bench Live       & 76.0 & 16.0 & 8.0 \\
\midrule
Overall (avg.)       & 78.0 & 12.0 & 10.0 \\
\bottomrule
\end{tabular}
\caption{Manual patch correctness of \eviAct{} with DeepSeek-V4-Pro (\%). Correct patches are semantically valid repairs, Overfitting patches pass the full test suite without implementing a general repair, and Unknown patches lack the evidence to decide.}
\label{tab:patch_correctness}
\end{table}

\subsection{Failure and Correctness Analysis (RQ3)}
\label{sec:failure-analysis}
Resolve rates show neither where repair still fails nor whether passing patches are correct. We therefore manually analyze 200 failed runs and 100 resolved patches of \eviAct{} with DeepSeek-V4-Pro, sampling 50 failed runs and 25 resolved patches per benchmark. Two authors annotate independently and resolve disagreements through discussion, reaching a Cohen's $\kappa$ of 0.84 for failure modes and 0.87 for patch correctness. Appendices~\ref{app:stratified_sampling}, \ref{app:failure-criteria}, and~\ref{app:patch_correctness_criteria} describe the sampling and annotation criteria.

Residual failures concentrate in localization and semantic editing. Table~\ref{tab:failure_modes} shows that mislocalization (45.0\%) and mis-edits (39.0\%) account for 84.0\% of the failures, and the share of mislocalization stays between 42.0\% and 48.0\% across benchmarks. Mislocalization occurs when graph expansion exhausts its budget before reaching a fix file that is structurally reachable from the RED evidence, or when the stack trace points to where the failure manifests instead of where the fault lies. Mis-edits retrieve the correct context but implement the wrong semantic change, which the GREEN target tests reject. Partial repairs (13.5\%, between 12.0\% and 14.0\% on every benchmark) pass the target tests but fail full regression, and timeouts (2.5\%) are rare. Mis-edits are most frequent on Verified (42.0\%) and timeouts on Lite (4.0\%), while the four failure modes keep the same order on every benchmark. None of the 200 sampled failures ends in a malformed or non-compilable patch, and every failure falls into one of the four modes in Table~\ref{tab:failure_modes}.

Most resolved patches are semantically correct. Table~\ref{tab:patch_correctness} shows that 78 of the 100 audited patches are correct, 12 overfit the tests, and 10 cannot be adjudicated, with the share of correct patches between 76.0\% and 80.0\% on every benchmark. Overfitting is most common on Live, at 16.0\% against 8.0--12.0\% elsewhere. Overfitting patches hard-code observed behavior, weaken program logic, or cover only the exercised branch. Unknown patches differ from the developer fix in ways that the issue and tests cannot resolve.
\section{Conclusion}
\label{sec:conclusion}

In this paper, we presented \eviAct{}, an evidence-driven agentic APR framework that turns execution evidence into repair decisions through three guardrails. Across four benchmarks, \eviAct{} achieves higher resolve rates than the strongest published GPT-4o baselines at lower per-instance API cost, and under a matched backbone and budget it outperforms mini-SWE-agent and Agentless. Ablations and manual analysis trace these gains to complementary guardrails, which ground the repair context, filter invalid patches, and leave mainly semantic failures. Enforcing the same checks adds 8.3 points over advisory feedback, so the gain comes from letting evidence decide the next action. Future work will extend this design toward recursive repair that feeds each attempt's execution feedback back into localization, editing, and validation.

\section*{Limitations}
\label{sec:limitations}


Benchmark contamination cannot be ruled out, since Defects4J~2.0, SWE-bench Verified, and SWE-bench Lite are public and the training data of the evaluated LLMs cannot be audited. SWE-bench Live draws on recent issues to reduce this risk, and \eviAct{} keeps its margin there. The reruns on a shared backbone also keep model knowledge identical across the compared frameworks.

Our evaluation covers Defects4J~2.0 and SWE-bench, which exercise Java and Python repository repair. These benchmarks span different project structures, issue types, and validation settings, but they leave out ecosystems such as C/C++, JavaScript/TypeScript, Rust, Go, mobile projects, and large polyglot systems. \eviAct relies on repository indexing, build feedback, target-test execution, and full validation, and ecosystems with weaker test infrastructure, non-standard builds, or flaky tests require additional engineering. Appendix~\ref{app:open_weight_models} reports results with two open-weight Qwen models of different sizes.

\looseness=-1 Agentless and mini-SWE-agent are rerun with DeepSeek-V4-Pro under matched instances and budgets. Published GPT-4o baselines provide context and are not controlled head-to-head experiments. On Defects4J~2.0, we could not reproduce RepairAgent and AdverIntent in our environment, so we report their published results, and the controlled comparison covers only the three SWE-bench sets. API cost also depends on provider pricing, so we report runtime and token use separately.

The retrieval scaffold ranks suspects by deterministic evidence alignment. It works best when failure evidence names concrete frames, symbols, or assertions. Some configuration-driven, data-dependent, or semantically distant root causes lie outside its bounded graph expansion, as the mislocalization share in Table~\ref{tab:failure_modes} reflects. Learned or adaptive retrieval fits the same guardrail interface.

The TD gate admits patches that recover the target tests, which in some cases favors test-specific fixes. Full regression reduces this risk without removing it, and 12.0\% of audited resolved patches overfit (Table~\ref{tab:patch_correctness}). The gates therefore complement semantic validation and do not replace it.

\bibliography{sample-base}

\appendix

\section{Default Configuration Rationale}
\label{app:config_rationale}

This appendix explains the default localization and budget settings used in \eviAct{}. They define a compact and reproducible operating point that bounds repository exploration, patch refinement, and validation cost across benchmarks.

\subsection{Suspect-Set Size}

\eviAct{} uses $K=3$ suspect spans as the default localized working set. This choice reflects a precision--recall trade-off in repository-level repair. A smaller suspect set gives the patch agent a compact context and reduces the risk of editing loosely related code, but it may miss the repair location when RED evidence points only to a failure manifestation. A larger suspect set increases the chance of including the correct repair context, but it also dilutes the working set and invites edits across weakly related files.

We therefore use $K=3$ as a compact default. It allows the retrieval scaffold to expose multiple structurally related candidates while keeping patch generation anchored to a small evidence-backed context. This setting also aligns with the intended role of the retrieval scaffold: to constrain the patch agent to a focused suspect set, not to provide a broad repository summary.

\subsection{Ranking Order}

The lexicographic ranking key prioritizes direct executable evidence before weaker structural signals. Exact trace-symbol matches are ranked first because stack frames, failing test names, and assertion-related symbols provide the strongest anchors when they are available. File or line proximity is ranked next because failure logs often identify the manifestation site even when they do not identify the repair site. Graph distance and relation support are then used to recover structurally connected contexts through containment, call, import, and inheritance relations. Span length is used only as a final tie-breaker, preferring compact spans when other evidence signals are equivalent.

This ordering is intentionally conservative. It favors candidates that are directly grounded in RED evidence before expanding to more indirect graph-neighborhood evidence. The design reduces unconstrained keyword search while still allowing the retrieval scaffold to move beyond the immediate failure frame when the repair location is structurally related but absent from the RED log.

\subsection{Tool-Call Budget}

\eviAct{} uses a fixed 56-call budget per instance, divided into 36 localization calls and 20 patching calls. The purpose of this budget is to make runs reproducible and to prevent long open-ended agent trajectories. The allocation gives more budget to localization because early repair drift is difficult to recover from. If the suspect set is wrong, additional patch attempts refine an incorrect repair hypothesis and leave the localization error in place.

The patching budget is kept bounded because candidate edits are filtered by the compile gate and then by the TD gate. In other words, \eviAct{} is designed to improve patch quality by controlling the evidence flow into patch generation, rather than by allowing many unconstrained patch retries. This budget allocation is therefore part of the framework's stage-boundary design: localization is allowed enough interaction to establish an evidence-backed working set, while patching is constrained to produce executable and target-recovering edits within that working set.

\section{Prompt Templates}
\label{app:prompts}

This appendix reports the prompt templates used by \eviAct{}. Placeholders such as \texttt{<BENCHMARK>}, \texttt{<LANGUAGE>}, and \texttt{<RED\_LOG\_PATH>} are filled automatically at runtime. The prompts enforce the same gated workflow used in the main method: RED evidence is first used for localization, candidate edits are then checked by the compile gate, and only compilable patches proceed to GREEN target-test validation and full regression.

\subsection{System Prompt}

\begin{promptbox}{System Prompt}
You are an agentic program repair system for <BENCHMARK>.

Global rules:
- Do not modify, skip, delete, or disable tests.
- Do not hardcode behavior only for the observed failing test.
- Produce minimal and localized patches.
- Preserve the syntax, style, and idioms of <LANGUAGE>.
- Prefer changes that explain the RED failure rather than broad rewrites.

Workflow:
1. Setup: check out the buggy revision and prepare the environment.
2. RED: run the originally failing target test and collect failure evidence.
3. Localize: use RED evidence and the repository index to identify suspicious files, symbols, and line ranges.
4. Patch: generate one minimal candidate edit over the localized working set.
5. Compile: apply the edit and run syntax or build checks.
6. GREEN: rerun the originally failing target test and require it to pass.
7. Validate: run full regression validation before accepting the patch.

When asked to localize, output only the requested localization JSON.
When asked to patch, output only the requested edit JSON or unified diff.
Do not include explanations unless explicitly requested by the controller.
\end{promptbox}
\subsection{Localization Prompt}

\begin{promptbox}{Localization Prompt}
Goal: localize the bug using RED failure evidence and the repository index.

Inputs:
- Workdir: <WORKDIR>
- RED log: <RED_LOG_PATH>
- Repository index: <INDEX_PATH>
- Target language: <LANGUAGE>

Instructions:
1. Read the RED log first.
2. Extract the failing test, assertion or exception message, and relevant stack frames.
3. Derive primary symbols from stack frames, test names, assertion text, and error messages.
4. Query the repository index using symbol lookup before textual search.
5. Read matched AST spans before selecting suspects.
6. Inspect caller, callee, import, inheritance, or reference spans only when they are structurally connected to the RED evidence.
7. Keep the working set small and evidence-grounded.

Output only valid JSON in the following schema:
{
  "red_test": "name of the failing target test",
  "red_failure_summary": "one-sentence failure summary",
  "failure_contract": [
    "behavior that should hold but is violated"
  ],
  "primary_symbols": [
    "symbol names derived from RED evidence"
  ],
  "suspects": [
    {
      "file": "relative/path/to/file",
      "symbol": "class, method, function, or field name",
      "start_line": 1,
      "end_line": 20,
      "evidence": "why this span is connected to the RED failure"
    }
  ],
  "working_set": [
    {
      "file": "relative/path/to/file",
      "symbol": "symbol to inspect or edit"
    }
  ]
}
\end{promptbox}

\subsection{Patch Prompt}

\begin{promptbox}{Patch Prompt}
Goal: fix the localized bug with a minimal candidate edit.

Inputs:
- RED evidence: <RED_EVIDENCE>
- Suspect spans: <SUSPECT_SPANS>
- Diagnostic context: <DIAGNOSTIC_CONTEXT>
- Target language: <LANGUAGE>

Instructions:
1. Patch only files that are supported by the suspect spans unless additional context is necessary.
2. Preserve unrelated behavior and avoid broad rewrites.
3. Do not modify tests.
4. Do not hardcode behavior only for the observed failing test.
5. Ensure that the edit applies cleanly and preserves syntax.
6. If the previous candidate failed to apply, compile, or pass GREEN, use the diagnostic context to revise only the relevant part of the patch.

Output only valid JSON in the following schema:
[
  {
    "path": "relative/path/to/file",
    "ops": [
      {
        "type": "replace",
        "start_line": 10,
        "end_line": 12,
        "text": "replacement code with preserved indentation\n"
      }
    ]
  }
]

The controller also accepts a unified diff if structured edits cannot express the change. In that case, output only the unified diff beginning with "diff --git".
Do not use markdown fences or explanations.
\end{promptbox}
\section{Positioning Against Prior Systems}
\label{app:positioning}

\looseness=-1 Table~\ref{tab:positioning} compares how agentic repair systems use execution evidence in the three stages that \eviAct{} guards, namely localization, patch, and validation. Prior systems already compute related signals: AutoCodeRover can consult spectrum-based fault localization, SWE-agent lints edits, Agentless applies, reproduces, and regression-tests candidate patches, and TDFlow organizes development around RED/GREEN tests. In these systems, the signal is either context that the model may use or a filter that removes or reranks candidates within a stage. In \eviAct{}, the outcome of each check decides whether the next stage runs and what it acts on. In terms of the triple $(e, c, a)$ of Section~\ref{sec:method}, a guardrail sets a constraint $a$ on the next stage, a filter applies $c$ only within a stage, and advisory evidence has no constraint $a$.

\begin{table}[!htb]
\centering
\small
\setlength{\tabcolsep}{4pt}
\renewcommand{\arraystretch}{1.1}
\begin{tabular*}{\columnwidth}{@{\extracolsep{\fill}}lccc@{}}
\toprule
\textbf{System} & \textbf{Localization} & \textbf{Patch} & \textbf{Validation} \\
\midrule
SWE-agent & \gnone & \ghalf & \gnone \\
OpenHands & \gnone & \gnone & \gnone \\
mini-SWE-agent & \gnone & \gnone & \gnone \\
AutoCodeRover & \gnone & \ghalf & \ghalf \\
Agentless & \gnone & \ghalf & \ghalf \\
TDFlow & -- & -- & \gnone \\
\eviAct{} (ours) & \gfull & \gfull & \gfull \\
\bottomrule
\end{tabular*}
\caption{Use of execution evidence in the localization, patch, and validation stages of SWE-agent~\citep{yang2024swe-agent}, OpenHands~\citep{wang2024openhands}, mini-SWE-agent~\citep{miniswe2025}, AutoCodeRover~\citep{zhang2024autocoderover}, Agentless~\citep{xia2024agentless}, TDFlow~\citep{lee2025tdflow}, and \eviAct{}. \gfull{}~marks a guardrail, whose check decides whether and on what the next stage runs. \ghalf{}~marks a filter, which removes or reranks candidates within a stage. \gnone{}~marks advisory evidence that the model is free to ignore, and -- marks a stage the system does not check.}
\label{tab:positioning}
\end{table}

\section{Evaluation Details}
\label{app:metrics}

Table~\ref{tab:metrics_app} defines the effectiveness and efficiency metrics used throughout the paper.
We use file-level Hit@3 because the patch agent operates over a small top-$K$ suspect set, and repository-level repair typically requires surfacing at least one file modified by the benchmark reference patch before finer-grained editing.

The TD gate operates on the target tests of each instance: the benchmark-provided triggering tests on Defects4J~2.0, and on SWE-bench, reproduction tests generated by the LLM from the issue description following Agentless~\citep{xia2024agentless}.
In Setup, \eviAct{} executes them on the original buggy revision to obtain RED failure evidence; in Verify, it reruns the same tests on the patched workspace to check whether the observed failure turns GREEN.
Final validation follows the official benchmark protocol.

\begin{table}[!htb]
  \centering
  \small
  \caption{Definitions of evaluation metrics for assessing the effectiveness of \eviAct and its efficiency.}
  \label{tab:metrics_app}
  \setlength{\tabcolsep}{4pt}
  \renewcommand{\arraystretch}{1.1}
  \begin{tabularx}{\linewidth}{@{}p{0.26\linewidth}X@{}}
    \toprule
    \textbf{Metric} & \textbf{Definition} \\
    \midrule
    \multicolumn{2}{c}{\textbf{Effectiveness}} \\
    \midrule
    Resolve (\%) & Percentage of instances that pass the benchmark validation suite. \\
    Hit@3 (\%) & Standard Top-3 file-level localization accuracy; an instance is counted as hit if any reference-modified file appears in the top three localized files. \\
    CompFail & Number of attempted patches rejected by the compile gate, including malformed diffs, failed hunk application, syntax errors, and build failures. \\
    ValFail & Number of attempted patches that compile but fail the target tests or full regression. \\
    \midrule
    \multicolumn{2}{c}{\textbf{Efficiency}} \\
    \midrule
    Cost (\$) & Average API cost per instance computed from token usage and vendor pricing. \\
    Runtime (s) & End-to-end wall-clock time per instance from localization to verification. \\
    Tokens (K) & Total LLM tokens consumed per instance, including input and output tokens, also reported by stage. \\
    Tool Calls & Repository, search, edit, build, or test invocations per instance, also reported by stage. \\
    \bottomrule
  \end{tabularx}
\end{table}

\section{Open-weight Model Results}
\label{app:open_weight_models}

To examine whether \eviAct can operate with open-weight backbones, Table~\ref{tab:open_weight_models} reports results with two Qwen variants \cite{qwen2024qwen25} on Defects4J~2.0 and SWE-bench Verified. 

\begin{table}[!htb]
\centering
\small
\caption{Resolve rates (\%) of \eviAct{} with Qwen.}
\label{tab:open_weight_models}
\setlength{\tabcolsep}{4pt}
\renewcommand{\arraystretch}{1.08}
\begin{tabularx}{\linewidth}{@{}Xcc@{}}
\toprule
Model & Defects4J~2.0 & SWE-bench Verified \\
\midrule
Qwen2.5-Coder-32B & 14.3\% & 15.2\% \\
Qwen2.5-72B & 18.0\% & 20.5\% \\
\bottomrule
\end{tabularx}
\end{table}

\begin{table}[!htb]
\centering
\footnotesize
\setlength{\tabcolsep}{3.5pt}
\renewcommand{\arraystretch}{1.12}
\begin{tabularx}{\linewidth}{@{}>{\raggedright\arraybackslash}p{0.28\linewidth}X@{}}
\toprule
\textbf{Failure mode} & \textbf{Criterion} \\
\midrule
\makecell[l]{Mislocalization\\{\scriptsize Localize}}
& The correct fix file or symbol is absent from the retrieved suspect set, so the agent lacks the necessary repair context. \\

\addlinespace[2pt]
\makecell[l]{Mis-edit\\{\scriptsize Patch}}
& The correct fix file or symbol is retrieved, but the executable patch modifies the wrong location or implements an incorrect semantic change, causing the GREEN target tests to fail. \\

\addlinespace[2pt]
\makecell[l]{Partial repair\\{\scriptsize Full regression}}
& The patch passes the GREEN target tests but fails full regression, indicating an incomplete repair or a missed edge case. \\

\addlinespace[2pt]
\makecell[l]{Timeout\\{\scriptsize Any stage}}
& The run exhausts the time or tool budget before reaching a stable validation outcome, including patch generation, GREEN test, or full regression. \\
\bottomrule
\end{tabularx}
\caption{Root-cause criteria used in failure-mode annotation. The stage shown under each label indicates where the failure is primarily assigned.}
\label{tab:failure_mode_criteria}
\end{table}

\begin{table}[!htb]
\centering
\footnotesize
\setlength{\tabcolsep}{3.5pt}
\renewcommand{\arraystretch}{1.12}
\begin{tabularx}{\linewidth}{@{}>{\raggedright\arraybackslash}p{0.24\linewidth}X@{}}
\toprule
\textbf{Label} & \textbf{Criterion} \\
\midrule
Correct
& The patch is semantically consistent with the developer fix or satisfies the issue requirement without weakening existing behavior. Syntactic differences from the developer patch are allowed. \\

\addlinespace[2pt]
Overfitting
& The patch passes benchmark validation but does not implement a general repair, e.g., by hard-coding observed behavior, weakening logic, or covering only the exercised branch or input pattern. \\

\addlinespace[2pt]
Unknown
& The patch differs from the developer fix or follows a different repair strategy, but the issue description, tests, and local context are insufficient to determine whether it is incorrect or a valid alternative repair. \\
\bottomrule
\end{tabularx}
\caption{Manual patch correctness criteria.}
\label{tab:patch_correctness_criteria}
\end{table}

On Defects4J~2.0, Qwen2.5-72B-Instruct reaches 18.0\%, below \eviAct with proprietary backbones and 3.6--4.3 pp below AdverIntent and RepairAgent, which use GPT-4o and GPT-3.5-Turbo.
This suggests that \eviAct can remain competitive on structured Java repair settings even with an open-weight backbone, especially when the benchmark provides executable failures and relatively stable repair environments.
On SWE-bench Verified, however, the Qwen variants remain behind GPT-4o baselines, indicating that repository-scale issue repair places higher demands on localization, tool use, and semantic patch generation.
Overall, these results show that \eviAct is compatible with open-weight models, but its effectiveness remains sensitive to the underlying model's ability to follow retrieved evidence and produce executable, semantically appropriate edits.

\section{Ablation and Sampling Protocols}
\label{app:stratified_sampling}

For failure-mode and patch correctness analyses, we use benchmark-wise stratified sampling to ensure coverage over project or repository groups.
For Defects4J~2.0, strata are defined by project.
For SWE-bench Verified, SWE-bench Lite, and SWE-bench Live, strata are defined by repository.
Sampling is performed without replacement.
The sampled instance IDs are released with the artifact.

For a target sample size $N$ on a dataset, we first assign each non-empty stratum at least one instance whenever possible.
The remaining quota is allocated approximately proportional to stratum size.
If a stratum contains fewer eligible instances than its allocated quota, all eligible instances from that stratum are included and the remaining quota is redistributed to larger strata.

For the component ablation, we use the complete SWE-bench Lite test set of 300 instances, without instance sampling.
All five configurations use the same 300 instances with the same backbone and budget: no guardrails, +Retrieval, +Compile, +TD, and full \eviAct.
This paired design keeps instance composition identical across configurations. Each configuration is run three times; Table~\ref{tab:ablation_seeds} reports the resolve rates.

\begin{table}[!htb]
\centering
\small
\setlength{\tabcolsep}{4pt}
\begin{tabular}{@{}lcc@{}}
\toprule
Configuration & Resolve (\%) & $\Delta$ (pp) \\
\midrule
No guardrails & 68.3$\pm$1.0 & -- \\
+Retrieval & 70.3$\pm$1.3 & +2.0 \\
+Compile & 75.5$\pm$2.0 & +7.2 \\
+TD & 75.2$\pm$1.0 & +6.8 \\
\eviAct{} & 81.0$\pm$2.0 & +12.7 \\
\bottomrule
\end{tabular}
\caption{Resolve rates of the guardrail ablation on SWE-bench Lite (300 instances) with DeepSeek-V4-Pro, mean$\pm$standard deviation over three runs. $\Delta$: gain over no guardrails.}
\label{tab:ablation_seeds}
\end{table}

For failure-mode analysis, we sample failed runs only.
We draw 50 failed instances from each benchmark under the DeepSeek-V4-Pro setting, yielding 200 failed runs in total.
The sampling follows the same project/repository stratification described above.
When a stratum has fewer failed runs than its allocated quota, all failed runs from that stratum are included and the residual quota is redistributed.

\looseness=-1 For patch correctness analysis, we sample resolved instances only.
We draw 25 full-validation-passing patches from each benchmark under the DeepSeek-V4-Pro setting, yielding 100 patches in total.
The same stratification and redistribution rules are applied to preserve project or repository coverage.

\section{Diagnostic Case Study}
\label{app:case}

Table~\ref{tab:case_study} illustrates the evidence chain on JacksonDatabind-18. The baseline follows a generic exception search and repeatedly edits an unrelated file. \eviAct uses the failure evidence to retrieve the relevant configuration path, applies a single patch, and validates it through the compile and TD gates.

\section{Failure Mode Annotation Criteria}
\label{app:failure-criteria}

For failure analysis, each failed run is assigned a primary failure label at the earliest stage where the evidence-to-action chain breaks. 
Annotators inspect the retrieved candidates, edited files, generated diffs, developer/reference fix when available, Compile-gate outputs, GREEN-test logs, full-regression logs, and timeout/tool records. 
When later failures are downstream effects of an earlier error, we label the earliest traceable cause.
Table ~\ref{tab:failure_mode_criteria} defines the labeling criteria for four failure modes.

\section{Patch Correctness Annotation Criteria}
\label{app:patch_correctness_criteria}
\looseness=-1 Patch correctness is annotated only for resolved instances. For each sampled patch, annotators inspect the generated diff, issue description, original failure log, full-validation result, and developer/reference fix when available.
The goal is to distinguish semantically valid repairs from patches that pass the benchmark tests but do not implement a general fix.
Table~\ref{tab:patch_correctness_criteria} defines the  annotation labels of patch correctness.

\section{Additional Experimental Comparisons}
\label{app:additional_comparisons}

\subsection{Baseline Rerun Configuration}
\label{app:baseline_reruns}
We rerun Agentless and mini-SWE-agent with DeepSeek-V4-Pro on SWE-bench Verified (500 instances), Lite (300 instances), and the same Live subset (100 instances) used for \eviAct. The reruns use the same model, per-instance budget, and execution environment as \eviAct; each baseline keeps its own repair workflow and default SWE-bench configuration. Results are averaged over three runs. Resolve rate follows the benchmark validation suite. API cost, total input/output tokens, and runtime are averaged over all attempted instances, including unsuccessful runs. Table~\ref{tab:rerun_efficiency} compares end-to-end resource use on Lite. \eviAct achieves both the lowest token use and the shortest runtime.

\begin{table}[!htb]
\centering
\small
\setlength{\tabcolsep}{5pt}
\renewcommand{\arraystretch}{1.08}
\begin{tabularx}{\linewidth}{@{}Xrrrr@{}}
\toprule
\textbf{Framework} & \shortstack{Resolve\\(\%)} & \shortstack{Cost\\(\$)} & \shortstack{Tokens\\(K)} & \shortstack{Runtime\\(s)} \\
\midrule
Agentless$^{*}$ & 42.0 & 0.13 & 1549.0 & 1943.5 \\
mini-SWE-agent$^{*}$ & 66.3 & 0.06 & 1312.0 & 433.6 \\
\eviAct & \textbf{81.0} & \textbf{0.01} & \textbf{145.0} & \textbf{348.5} \\
\bottomrule
\end{tabularx}
\caption{Repair effectiveness and end-to-end resource use on SWE-bench Lite (300 instances) with DeepSeek-V4-Pro. Values are per-instance means over all attempted instances. $^{*}$ marks baseline reruns.}
\label{tab:rerun_efficiency}
\end{table}

Table~\ref{tab:rerun_significance} reports run-to-run variance and paired significance. Confidence intervals come from a paired bootstrap over instances with 10{,}000 resamples. For McNemar's test, we aggregate the three runs by majority voting, counting an instance as resolved if at least two of the three runs succeed, and then apply a two-sided exact McNemar test to the paired instance-level outcomes of \eviAct{} and each baseline, separately for each dataset.

\begin{table}[!htb]
\centering
\footnotesize
\setlength{\tabcolsep}{2pt}
\renewcommand{\arraystretch}{1.08}
\resizebox{\linewidth}{!}{%
\begin{tabular}{@{}l l r r c r@{}}
\toprule
\textbf{Set} & \textbf{Method} & \textbf{Mean$\pm$Std} & \textbf{$\Delta$} & \textbf{95\% CI} & \textbf{$p$} \\
\midrule
Verified & \eviAct & 75.6$\pm$1.4 & -- & -- & -- \\
 & mini-SWE-agent & 65.6$\pm$1.6 & +10.0 & [7.27, 12.73] & $<$0.001 \\
 & Agentless & 48.8$\pm$1.6 & +26.8 & [23.80, 29.73] & $<$0.001 \\
\midrule
Lite & \eviAct & 81.0$\pm$2.0 & -- & -- & -- \\
 & mini-SWE-agent & 66.3$\pm$2.0 & +14.7 & [11.00, 18.44] & $<$0.001 \\
 & Agentless & 42.0$\pm$2.0 & +39.0 & [35.11, 42.89] & $<$0.001 \\
\midrule
Live & \eviAct & 44.0$\pm$4.0 & -- & -- & -- \\
 & mini-SWE-agent & 25.0$\pm$3.0 & +19.0 & [13.33, 24.67] & $<$0.001 \\
 & Agentless & 9.0$\pm$2.0 & +35.0 & [29.67, 40.33] & $<$0.001 \\
\bottomrule
\end{tabular}}
\caption{\looseness=-1 Run-to-run variance and paired significance of the DeepSeek-V4-Pro reruns (three runs). $\Delta$: resolve-rate gain of \eviAct{} over the baseline in percentage points.}
\label{tab:rerun_significance}
\end{table}

\subsection{Enforced versus Advisory Guardrails}
\label{app:advisory}
The advisory variant uses the same backbone (DeepSeek-V4-Pro), instances (all 300 SWE-bench Lite instances), prompts, and budgets as full \eviAct{}. Each guardrail check is still executed: the retrieval scaffold produces its suspect set, the compile gate produces diff, apply, and build diagnostics, and the TD gate produces GREEN outcomes. In the advisory variant, these outcomes are appended to the agent's context as feedback, but no transition is blocked. Both variants are run with three runs. Table~\ref{tab:advisory} reports means $\pm$ standard deviations. Under majority voting over runs, 42 instances are resolved only with enforced guardrails and 11 only with advisory guardrails (two-sided exact McNemar $p=2.25\times10^{-5}$); the paired bootstrap 95\% confidence interval of the gain is [5.67, 11.11] percentage points.

\begin{table}[!htb]
\centering
\small
\setlength{\tabcolsep}{5pt}
\renewcommand{\arraystretch}{1.08}
\begin{tabularx}{\linewidth}{@{}Xrrr@{}}
\toprule
\textbf{Variant} & \textbf{Resolve (\%)} & \textbf{Cost (\$)} & \textbf{$\Delta$ (pp)} \\
\midrule
Advisory & 72.7$\pm$1.3 & 0.034 & -- \\
Enforced (\eviAct{}) & \textbf{81.0}$\pm$2.0 & \textbf{0.01} & +8.3 \\
\bottomrule
\end{tabularx}
\caption{Enforced versus advisory guardrails on SWE-bench Lite (300 instances) with DeepSeek-V4-Pro, three runs. Cost is the mean per-instance API cost.}
\label{tab:advisory}
\end{table}

\subsection{Retrieval Backend Comparison}
\label{app:retrieval_comparison}
To examine the choice of retrieval backend, we compare BM25, SweRankEmbed-Small, SweRankEmbed-Large~\citep{reddy2026swerank}, and the \eviAct retrieval scaffold under the same DeepSeek-V4-Pro localization controller of full \eviAct{} on all 300 SWE-bench Lite instances. The scaffold's function-level Hit@3 of 93.0\% is therefore bounded by the file-level Hit@3 of full \eviAct{} (94.6\%), and the +Retrieval variant of Section~\ref{sec:ablation} runs without the other guardrails and is not directly comparable. Each retriever is used as in its original design, including its own query construction, and all operate on the pre-patch production code. The scaffold also takes the RED failure log as input, which BM25 and SweRankEmbed do not use, so the comparison measures evidence-driven retrieval as a whole. Hit@$k$ measures the proportion of instances in which a function modified by the reference patch occurs among the final top-$k$ functions. Figure~\ref{fig:retrieval} reports the results.

\begin{table*}[t]
\centering
\small
\caption{Diagnostic case study on JacksonDatabind-18 (Defects4J~2.0) comparing the Baseline and \eviAct execution traces across stages (Setup/Localize/Patch/Verify). \textcolor{red}{Red} highlights incorrect stack-trace hits, and \textcolor{blue}{blue} highlights correct stack-trace hits.}
\label{tab:case_study}
\begin{tabular}{>{\raggedright\arraybackslash}p{2cm}|>{\raggedright\arraybackslash}p{5.2cm}|>{\raggedright\arraybackslash}p{5.9cm}}
\toprule
\textbf{Stage} & \textbf{Baseline} & \textbf{\eviAct} \\
\midrule

Stack Trace & \multicolumn{2}{p{11.4cm}}{%
\vspace{-0.2em}%
\begin{minipage}{\linewidth}
\small
\colorbox{gray!10}{%
\begin{minipage}{\dimexpr\linewidth-2\fboxsep\relax}
\raggedright
\vspace{0.2em}
\codetext{\bluehl{junit.framework.AssertionFailedError}:} \newline
\codetext{expected:<sun.util.calendar.ZoneInfo[id="\bluehl{America/Los\_Angeles}",...]>} \newline
\codetext{but was:<sun.util.calendar.ZoneInfo[id="\bluehl{GMT}",...]>} \newline
\codetext{  at \bluehl{TestConfig.\bluehl{testDateFormatConfig}}(\bluehl{TestConfig.java:221})} \newline
\vspace{0.2em}
\end{minipage}%
}%
\end{minipage}%
\vspace{-0.2em}
} \\
\midrule

\multirow{2}{*}{1. Setup} 
& \textbf{Process:} Run test suite & \textbf{Process:} Run test suite \\
& \textbf{Output:} \texttt{test.trigger.log} & \textbf{Output:} \texttt{red.log} \gatered{[TD RED]} \\
\midrule

\multirow{3}{*}{2. Localize} 
& \textbf{Process:} \newline
  $\bullet$ Read log file \newline
  $\bullet$ Search for "Exception" keyword \newline
  $\bullet$ Read \redhl{MappingIterator.java} & \textbf{Process:} \newline
  $\bullet$ Read \texttt{red.log} \gatered{[TD RED]} \newline
  $\bullet$ Lookup \bluehl{ObjectMapper\#setDateFormat} \newline
  $\bullet$ Read span of \bluehl{BaseSettings.java} \gateblue{[Retrieval]} \\
& \textbf{Output:} \newline Located file: \redhl{MappingIterator.java} $\times$  & \textbf{Output:} \newline Located file: \bluehl{BaseSettings.java} $\checkmark$ \\
& API calls: 11, Tokens: 35.4K & API calls: 6, Tokens: 20.6K \\
\midrule

\multirow{6}{*}{3. Patch} 
& \textbf{Process:} \newline
  $\bullet$ Read \redhl{MappingIterator.java} \newline
  $\bullet$ Generate 1 attempt \newline
  $\bullet$ Generate 2 attempt \newline
  $\bullet$ Generate 3 attempt \newline
  $\bullet$ Generate 4 attempt \newline
  $\bullet$ ...
& \textbf{Process:} \newline
  $\bullet$ Read span of \bluehl{TestConfig.java} and \bluehl{BaseSettings.java} \gateblue{[Retrieval]} \newline
  $\bullet$ Generate 1 patch \newline
  $\bullet$ Compile check \gateorange{[Compile]} \newline
  $\bullet$ Verify with \texttt{green.log} \gategreen{[TD GREEN]} \\
& \textbf{Output:} 19 attempts & \textbf{Output:} 1 attempt, success \\
& API calls: 20, Tokens: 203.9K, Compile: $\emptyset$ & API calls: 4, Tokens: 30.3K, Compile: $\checkmark$ \\
\cmidrule{2-3}
& \multicolumn{1}{p{5.2cm}|}{%
\textbf{Patch Diff:} \newline
\vspace{-0.2em}%
\begin{minipage}{\linewidth}
\footnotesize
\ttfamily
\colorbox{gray!10}{%
\begin{minipage}{\dimexpr\linewidth-2\fboxsep\relax}
\raggedright
\vspace{0.2em}
\codetext{File: \redhl{MappingIterator.java}} \newline
\codetext{- if (t == null || t == JsonToken.END\_ARRAY) \{} \newline
\codetext{+ if (t == JsonToken.END\_ARRAY) \{ ... \}} \newline
\codetext{+ if (t == null) \{ ... \}} \newline
\codetext{.} \newline
\codetext{.} \newline
\codetext{.} \newline
\vspace{0.2em}
\end{minipage}%
}%
\end{minipage}%
\vspace{-0.2em}
} & \multicolumn{1}{p{5.9cm}}{%
\textbf{Patch Diff:} \newline
\vspace{-0.2em}%
\begin{minipage}{\linewidth}
\footnotesize
\ttfamily
\colorbox{gray!10}{%
\begin{minipage}{\dimexpr\linewidth-2\fboxsep\relax}
\raggedright
\vspace{0.2em}
\codetext{File: \bluehl{BaseSettings.java}} \newline
\codetext{- TimeZone tz = (df == null) ? \_timeZone : df.getTimeZone();} \newline
\codetext{+ // Should NOT change TimeZone when DateFormat changes} \newline
\codetext{  return new \bluehl{BaseSettings}(...,} \newline
\codetext{-     tz, \_defaultBase64);} \newline
\codetext{+     \_timeZone, \_defaultBase64);} \newline
\vspace{0.2em}
\end{minipage}%
}%
\end{minipage}%
\vspace{-0.2em}
} \\
\midrule

\multirow{3}{*}{4. Verify} 
& \textbf{Process:} \newline
  $\bullet$ Run full test suite \newline
  $\bullet$ 19 failures remain & \textbf{Process:} \newline
  $\bullet$ Run full test suite \newline
  $\bullet$ All tests passed \\
& \textbf{Output:} Result: Failed, \newline Time: 1129s, Total tokens: 239.3K & \textbf{Output:} Result: Success,\newline  Time: 153.6s, Total tokens: 50.9K \\
\bottomrule
\end{tabular}
\end{table*}

\end{document}